\documentclass[a4paper,11pt]{article}
\usepackage{pos}
\usepackage{graphicx}
\usepackage{subcaption}

\title{Studying SNR-MC interactions as galactic PeVatrons \\ in the era of CTAO and ASTRI Mini-Array}

\author*[a,b]{Alan Sunny}
\author[a]{Martina Cardillo}
\author[c]{Antonio Tutone}

\onbehalf{for ASTRI Mini-Array and CTAO Collaborations}

\affiliation[a]{Istituto di Astrofisicae Planetologia Spaziali- INAF,
  Via del Fosso del Cavaliere 100, 00133 Rome, Italy}

\affiliation[b]{Macroarea di Scienze MMFFNN, Università di Roma Tor Vergata,
Via della Ricerca Scientifica 1, 00133 Rome, Italy}

\affiliation[c]{Istituto di Astrofisica Spaziale e Fisica Cosmica -INAF,
Via Ugo La Malfa 153, I-90146 Palermo, Italy}

\emailAdd{alan.sunny@inaf.it}
\emailAdd{martina.cardillo@inaf.it} \emailAdd{antonio.tutone@inaf.it}

\abstract{Supernova remnants (SNRs) are widely recognized as key accelerators of Galactic cosmic rays (CRs), supported by the detection of the characteristic \textit{pion bump} in the $\gamma$-ray spectra of several SNRs. However, the recent observation of ultra-high-energy (UHE; $>$100 TeV) $\gamma$-rays by LHAASO from sources such as W51 region challenges standard models, which predict CR acceleration up to PeV energies only during the early ($\sim$100 year) phase of SNR evolution. Given the older age of known SNRs, alternative mechanisms — such as the interaction of runaway CRs with nearby molecular clouds (MCs) — have been proposed to explain the persistent UHE emission. In this study, we focus on the W51 complex, particularly the W51C-B region, as a promising site for investigating SNR–MC interactions. Simulated observations with the CTAO and the ASTRI Mini-Array are presented to demonstrate their crucial role in bridging the energy gap between Fermi-LAT and LHAASO, especially in the 0.3–100 TeV range. Their improved angular resolution will also help disentangle emission components from the interaction zone and nearby sources. Our theoretical modeling suggests that accelerated particles at the shock can account for the radio and GeV data, while UHE emission could be best explained by the combined contribution from both acceleration and adiabatic compression of cloud material at the SNR–MC interface.}

\ConferenceLogo{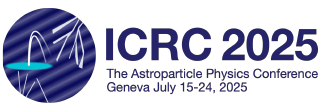}

\FullConference{39th International Cosmic Ray Conference (ICRC2025)\\
 15–24 July 2025\\
Geneva, Switzerland\\}

\begin{document}
\maketitle

\section{Introduction}
Supernova remnants (SNRs) are widely regarded as primary contributors to the population of Galactic cosmic rays (CRs), alongside other potential accelerators such as pulsar wind nebulae (PWNe), microquasars (MQs), and young massive star clusters (YMSCs). In the context of SNRs, particle acceleration is typically described by the diffusive shock acceleration (DSA) mechanism and observations of several SNRs in the $\gamma$-ray domain, especially the characteristic “pion bump” by the AGILE-GRID and Fermi-LAT  \cite{giuliani2011neutral, ackermann2013detection}, provide strong support for this process.  
A major breakthrough came with LHAASO’s detection of UHE $\gamma$-rays, revealing several Galactic PeVatrons, including SNRs \cite{lhaaso2021ultrahigh}. However, this challenges current theory that SNRs can can accelerate CRs to PeV energies only within their first $\sim$100 years-far younger than any known SNR \cite{cardillo2016supernova}.This apparent discrepancy has led to alternative scenarios, like SNRs indirectly interacting with nearby molecular clouds (MCs), illuminating them with accelerated CRs and leading to UHE $\gamma$-ray emission \cite{gabici2019origin}. This framework helps reconcile the age discrepancy and highlights SNR–MC interactions as prime environments for studying both the \textit{pion bump} and UHE signatures, making the identification of suitable observational targets essential. Here we present W51 as a compelling target. 

LHAASO recently published an important result on UHE emission from the W51 complex, reporting an unprecedented maximum energy limit of $\approx 200$ TeV \cite{cao2024evidence}. Located at the tangential point ($l = 49^\circ$) of the Sagittarius arm and about 4.5 kpc away, W51 spans nearly 100 pc in diameter and hosts two active star-forming regions, W51A to the north and W51B to the east, along with the middle-aged supernova remnant W51C (G49.2–0.7). The total mass of the complex is around $10^6 M_\odot$. Previous observations by Fermi-LAT revealed a prominent \textit{pion bump} in the $\gamma$-ray spectrum of W51C \cite{jogler2016revealing}, providing strong support for a hadronic origin of the lower-energy $\gamma$-ray emission. $\gamma$-rays in the energy range from tens of MeV to $\sim$30 TeV was also observed by H.E.S.S., Milagro, MAGIC, and HAWC \cite{abdalla2018hess, abdo2009milagro, aleksic2012morphological, fleischhack2019survey}. 
This dual signature - both \textit{pion bump} and $\sim$200 TeV photons—makes W51 an exceptional candidate for studying CR acceleration mechanisms.

The leading hypothesis for the origin of the emission is the interaction of accelerated CR protons from W51C with the nearby molecular clouds (MCs) in W51B \cite{fang2010non}. Other PeVatron candidates in the region include the PWNe candidate CXO J192318.5+140305 and several YMSCs \cite{cao2024evidence}. MAGIC detected a sub-TeV component ($\sim$20\% of the flux) likely from the PWNe, stressing the need for >1 TeV observations to resolve the morphology. LHAASO, due to its limited PSF, could not resolve the emission sites. Currently, the SNR–MC interaction model is best supported—backed by OH maser detections \cite{brogan2013oh} and evidence of a re-formed molecular clump at the interface \cite{tu2025molecular}. While it accounts for GeV to sub-5 TeV emission, it fails to explain the UHE emission observed by LHAASO. 

To investigate this discrepancy, we propose a combined acceleration + crushed cloud model: CRs are accelerated at the SNR shock, which also compresses nearby dense clouds, boosting magnetic fields and particle densities, making the cloud an efficient hadronic target \cite{blandford1982radio}.  CRs premeating the cloud can also be re-accelerated at the SNR shock enhancing emission further \cite{cardillo2016supernova}, but the process is not explored here. 
Our focus on the compression scenario is further motivated by LHAASO’s observation of a relatively larger particle population below 10 TeV compared to MAGIC. Although this discrepancy could be partially attributed to differences in detector sensitivity and angular resolution between extensive air shower arrays and imaging atmospheric Cherenkov telescopes (IACTs) \cite{cao2024evidence}, its underlying cause remains unresolved. HAWC has also reported a similar excess \cite{fleischhack2019survey}. If this model cannot explain the UHE component, an alternative scenario involves delayed $\gamma-$ray emission from CRs that escaped in the early SNR phase illuminating the MCs \cite{fang2010non}. 

In this context, the Cherenkov Telescope Array Observatory (CTAO) and the ASTRI Mini-Array will be key to resolving the region’s complex morphology. Their superior angular resolution and PSF will help distinguish between contributions from W51C, nearby PWNe, and other sources. They will provide critical coverage in the 0.3 - 200 TeV energy range, which is vital for probing the particle population below 10 TeV and for testing competing models. Taking advantage of the important results reported by the LHAASO collaboration, this work aims to investigate the potential of CTAO and ASTRI Mini-Array in studying this complex source morphology.

In Section \ref{Sec:SimuAnalysis}, we describe the simulation setup and highlight the importance of CTAO and ASTRI Mini-Array observations for resolving detailed emission features. Section \ref{Sec:Theory} introduces a theoretical framework to interpret the emission, and our conclusions are summarized in Section \ref{Sec:Conclusion}.

\vspace{-0.2cm}
\section{Simulation and Analysis}\label{Sec:SimuAnalysis}
The CTAO will be an observatory composed of tens of telescopes with three distinct sizes, strategically distributed across two sites: La Palma (Spain) in the Northern Hemisphere and Paranal (Chile) in the Southern Hemisphere. Designed to cover an extensive energy range from 20 GeV to beyond 300 TeV, CTAO will offer unprecedented sensitivity, angular resolution $(0.03^\circ$ above $10\, \text{TeV})$, and energy reconstruction capabilities \cite{mazin2019cherenkov}. The ASTRI Mini-Array \cite{scuderi2022astri} consists of nine small-size (4 m) Cherenkov telescopes currently under construction at the Teide Observatory in Tenerife. Expected to be completed within two years, it will offer unmatched sensitivity in the multi-TeV range and a good angular resolution of $0.05^\circ$ at $100\, \text{TeV}$ \cite{lombardi2022performance}, enabling detailed studies of candidate PeVatrons and their morphology through deep observations ($>$100 hrs per source).

To simulate and analyse the W51 region we use Gammapy v1.3 \cite{gammapy:2023}. The source is simulated within a $3^\circ \times 3^\circ$ field of view, centered on the position of W51C (RA = 290.830, Dec = 14.1558), using a spatial bin size of $0.02^\circ$. The analysis covers an energy range of 0.3–200 TeV for CTAO and 1.1–300 TeV for the ASTRI Mini-Array, divided into 20 logarithmically spaced energy bins. Instrument Response Functions (IRFs) corresponding to CTAO-North \cite{CTAO_2021_IRF} and ASTRI Mini-Array \cite{ASTRI_2022_IRF} are employed for simulation and analysis. These IRFs include effective area, angular and energy resolution, as well as the background rate and are generated for specific exposure times. The CTAO-North IRFs are chosen to ensure compatibility with the ASTRI Mini-Array’s Northern Hemisphere location, thereby facilitating a coherent comparative analysis. Observation times of 200 hours for CTAO and 300 hours for the ASTRI Mini-Array are considered: these are necessary not for the detection, since a $5\sigma$ significance is already reached within approximately 50 hours for both, but rather to improve the morphological resolution of the emission regions.

The maximum acceleration limit for protons may manifest in the $\gamma$-ray spectrum as a characteristic exponential cut-off feature \cite{cao2024evidence}. So, as LHAASO reported, we base our simulations on the exponential cut-off model they consider as the "expected total" model, with $A = 1.30 \times 10^{-15}\, \text{cm}^{-2}\, \text{s}^{-1}\, \text{TeV}^{-1}$, $\Gamma = 2.6$ and reference energy of 20 TeV. We also consider a cut off at 75 TeV:
\begin{align}
    \frac{dN}{dE} = A\, (E/20\, \text{TeV})^{-\Gamma} \, \text{exp}(-E/75\, \text{TeV})
\end{align}
Given CTAO’s expected capability to resolve the emission region and the potential of ASTRI Mini-Array to distinguish emission contributions from W51C and W51B, we model them as two separate sources using distinct spatial templates. For W51C, we adopt the Fermi-detected morphology as the spatial model, while for W51B, which lacks strong observational counterparts, we assume a uniform disk model with a $0.12^\circ$ extension. The objective is to fit the models of both sources such that their combined emission aligns with LHAASO’s observed spectrum. Residuals are computed over the defined region using the difference-to-sqrt(model) normalization, where the model represents the total emission from both W51C and W51B. Our goal is to minimize residual deviations and maintain consistency across the energy range. The current best fit spectral models we get are :
\begin{table}[h!]
\centering
\begin{tabular}{|c|c|}
\hline
\textbf{W51C} & \textbf{W51B} \\
\hline
\begin{tabular}[c]{@{}c@{}}
$\text{PLExpCut: } \displaystyle\frac{dN}{dE} = A\, \left( \frac{E}{20\, \text{TeV}}\right)^{-2.6} \exp\left( -\frac{E}{75\, \text{TeV}}\right)$ \\
$A = 8 \times 10^{-16}\, \text{cm}^{-2}\, \text{s}^{-1}\, \text{TeV}^{-1}$
\end{tabular}
&
\begin{tabular}[c]{@{}c@{}}
$\text{Power-law: } \displaystyle\frac{dN}{dE} = A_p \left( \frac{E}{20\, \text{TeV}} \right)^{-2.5}$\\
$A_p = 1.1 \times 10^{-16}\, \text{cm}^{-2}\, \text{s}^{-1}\, \text{TeV}^{-1}$
\end{tabular}
 \\
\hline
\end{tabular}
\end{table}

\begin{figure}[htbp]
    \centering
    \begin{subfigure}[b]{0.40\textwidth}
        \centering
        \includegraphics[width=\textwidth]{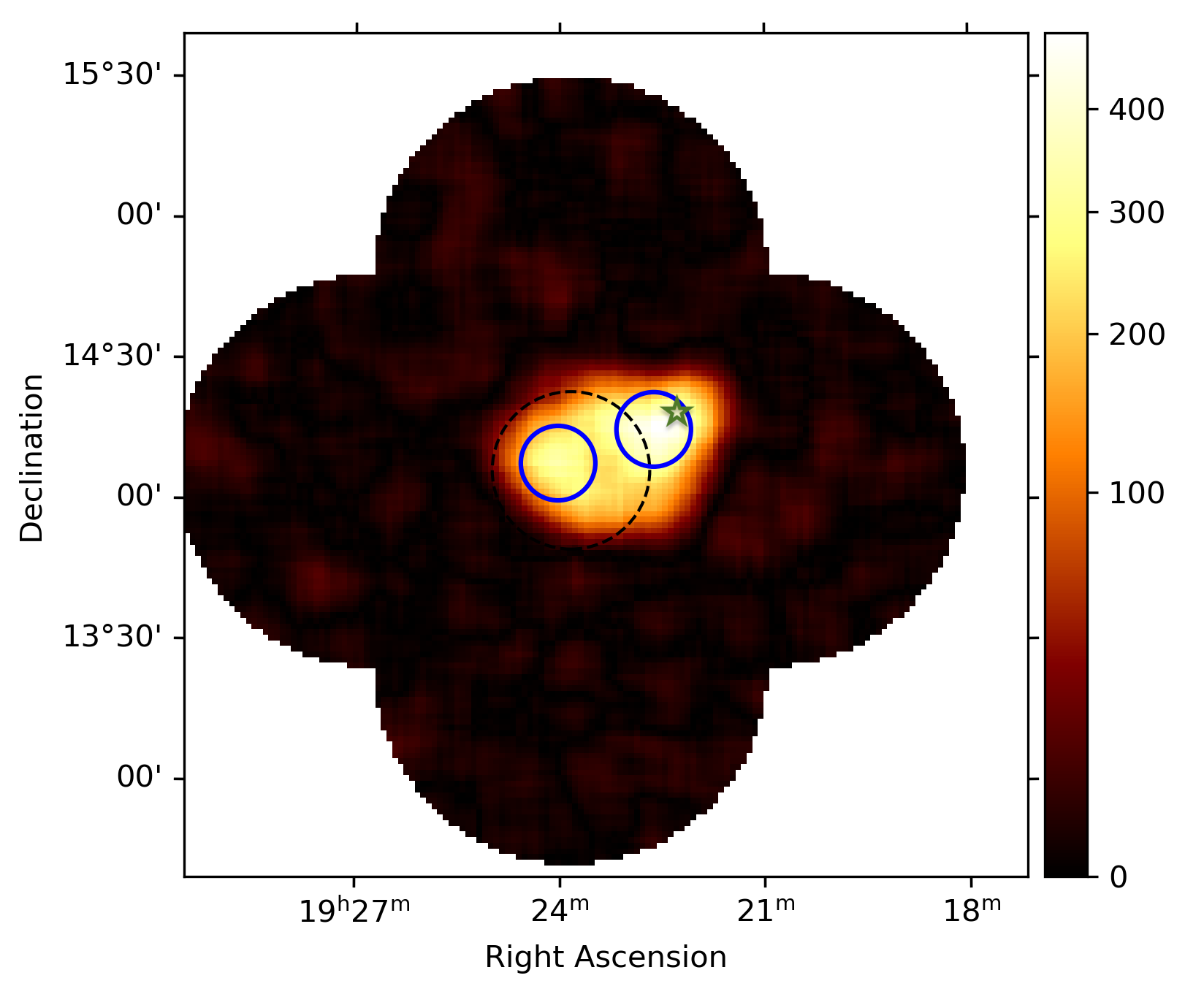}
        \caption{CTAO-North TS map}
        \label{fig:ctao-ts}
    \end{subfigure}
    \hfill
    \begin{subfigure}[b]{0.45\textwidth}
        \centering
        \includegraphics[width=\textwidth]{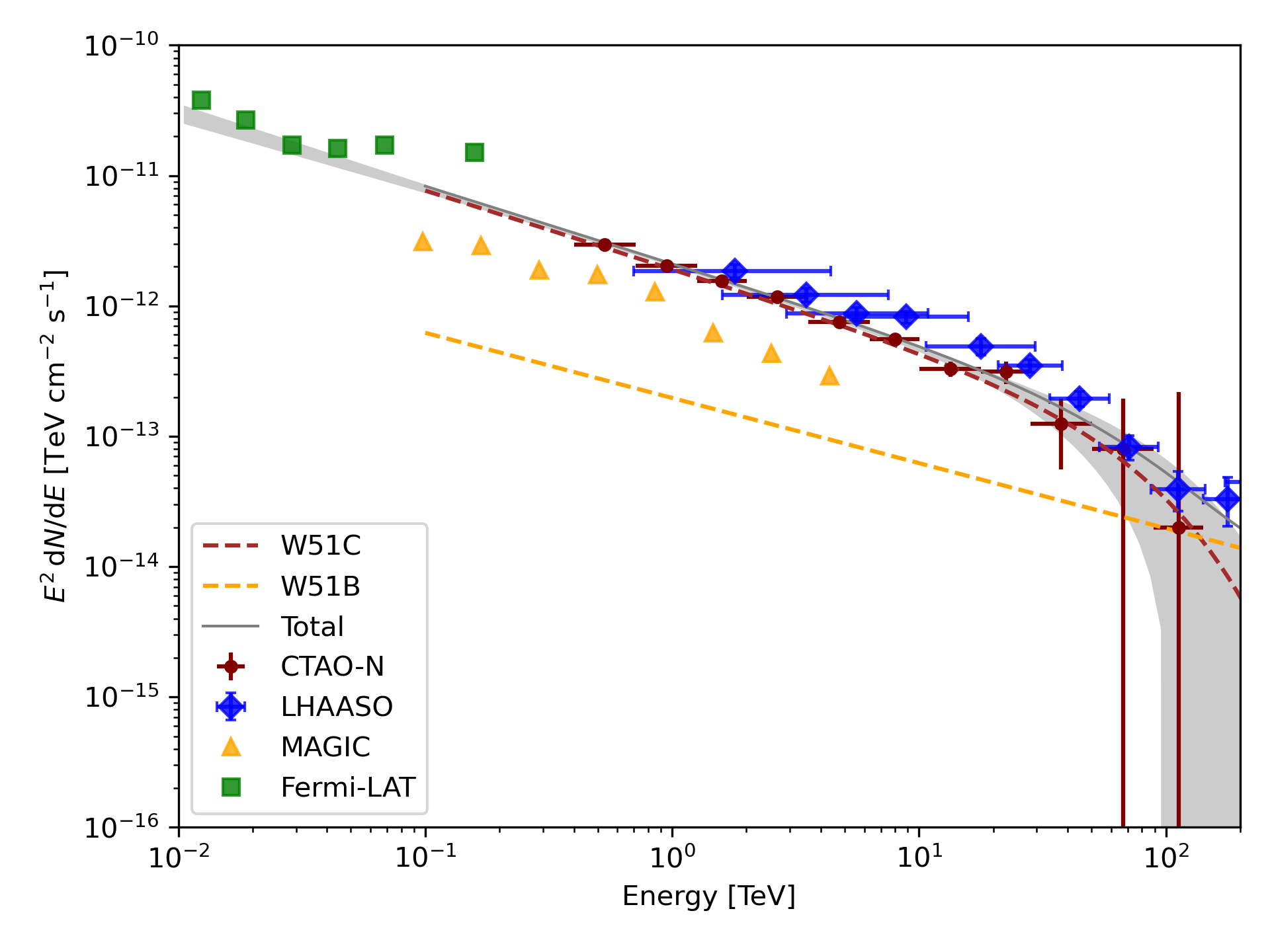}
        \caption{CTAO-North SED}
        \label{fig:ctao-sed}
    \end{subfigure}

    \vspace{0.5cm} 

    \begin{subfigure}[b]{0.40\textwidth}
        \centering
        \includegraphics[width=\textwidth]{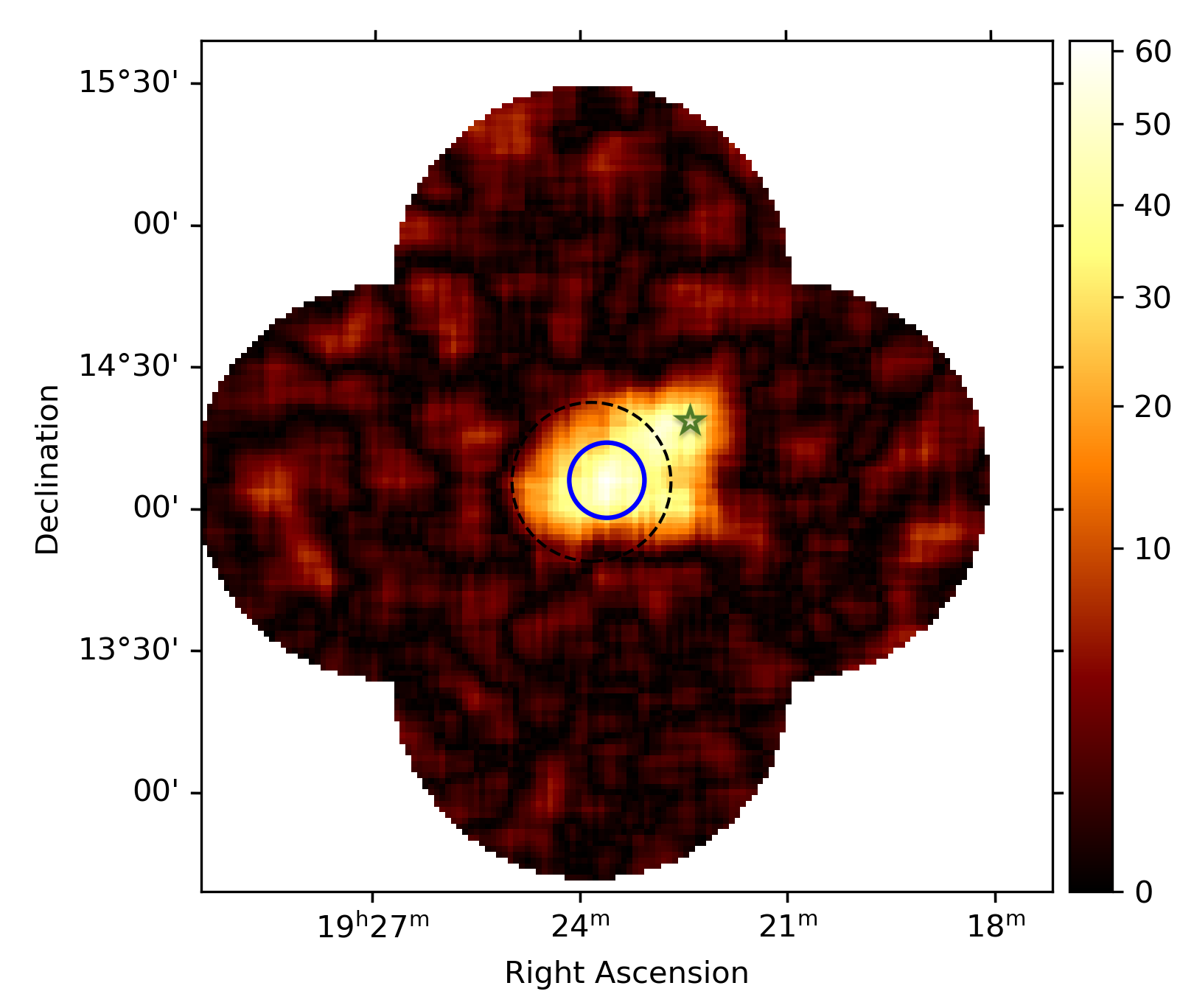}
        \caption{ASTRI Mini-Array TS map}
        \label{fig:astri-ts}
    \end{subfigure}
    \hfill
    \begin{subfigure}[b]{0.45\textwidth}
        \centering
        \includegraphics[width=\textwidth]{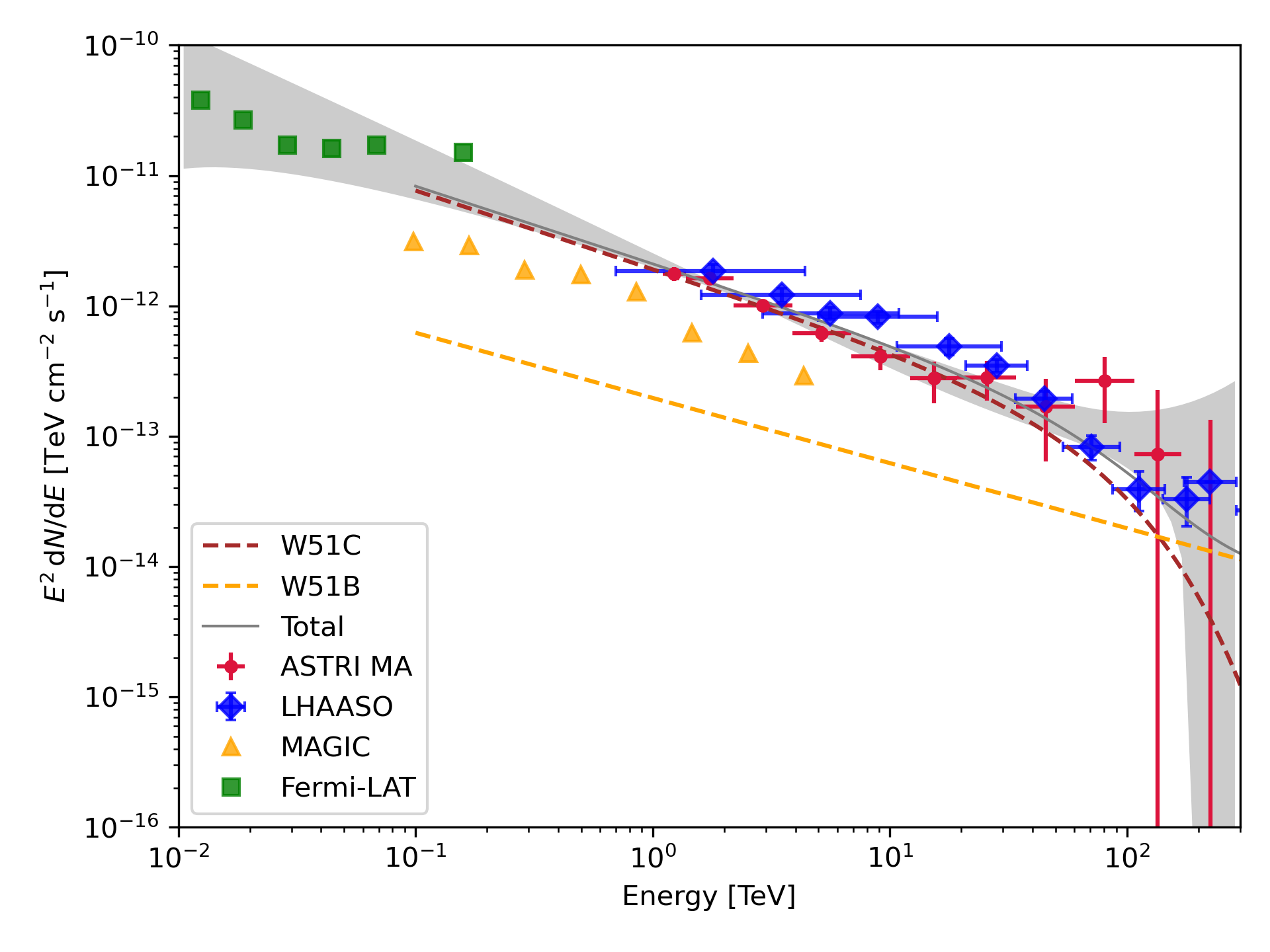}
        \caption{ASTRI Mini-Array SED}
        \label{fig:astri-sed}
    \end{subfigure}

    \caption{Detection and spectral modeling results for W51C-B with CTAO and ASTRI Mini-Array with 200 and 300 Hrs of exposure respectively. The LHAASO, Fermi-LAT and MAGIC data points are also included \cite{cao2024evidence, jogler2016revealing,aleksic2012morphological} The dashed circle shows the extension of the SNR and the green star represents the position of the W51B MC. The blue circle represents the points of $\geq 4 \sigma$ detection. (a) and (b) show the TS map and SED for CTAO-N, respectively; (c) and (d) show the same for the ASTRI Mini-Array.}
    \label{Fig:ctao-astri-results}
\end{figure}

We present the TS (test statistic) maps using a $4\sigma$ significance threshold (Figures \ref{fig:ctao-ts}, \ref{fig:astri-ts}). As anticipated, CTAO, with 200 Hrs exposure, detected two $4\sigma$ regions aligned with the SNR and the SNR–MC interaction region, while the ASTRI Mini-Array, with 300 Hrs of exposure, revealed a single detection within the SNR, near the interaction region. This difference is consistent with their respective angular resolution and energy thresholds.  Although we are not yet able to clearly separate the emission spectrum from the regions, improved modeling—particularly with ASTRI MA—may help achieve this.  Our findings are consistent with LHAASO results, indicating a potentially enhanced particle population below 10 TeV (refer to Figures \ref{fig:ctao-sed}, \ref{fig:astri-sed}), though further analysis is needed to confirm this. Although we obtain upper limits around 100 TeV—particularly for ASTRI Mini-Array —stronger constraints are expected with improved spatial modeling, currently under investigation. In the following section (Sec.\ref{Sec:Theory}), we explore a possible explanation for the ultra-high-energy emission and the enhanced particle population. The notable TS significance from the interaction region suggests an underlying mechanism that boosts the emission, which we propose as a combined effect of particle acceleration and adiabatic compression. 

\section{Physical aspects of the SNR-MC interaction scenario}\label{Sec:Theory}

Blandford et al. \cite{blandford1982radio} introduced the “crushed cloud” scenario, where a supernova-driven shock compresses a dense cloud, increasing the emission and explaining the radio observations. If the shock velocity and ambient density are sufficiently high, the post-shock gas becomes radiative, emitting ionizing radiation throughout the shock’s path. Initially, the gas is compressed, and as recombination sets in, further radiative cooling enhances compression and increases density. It is thus natural to explore whether the same mechanism contributes to $\gamma$-ray emission. 

For a shock velocity ($v_{\text{sh}}$) of $\sim 480$ km/s in W51C, the formation of a radiative zone requires a minimum column density $N_{\text{cool}} \approx 3 \times 10^{17} v_{\text{sh}}^4\, \text{cm}^{-2}$, leading to a minimum cloud density $n_{cloud} \gtrsim n_{cool}$. 
Also, the compression factor between the downstream of the shock (density $n_d$) and the crushed cloud (density $n_{cc}$), $s \equiv \frac{n_{cc}}{n_d}$ (or compression due to radiative cooling), can be limited by magnetic or thermal pressure. The compression due to radiative cooling enhances the particle energy spectrum normalization by a factor of $s^{2/3}$, and the momentum of each particle as $p \rightarrow p\,s^{1/3}$. 

\subsection{Contributions from acceleration and adiabatic compression of clouds}

The primary mechanism for $\gamma$-ray emission in SNRs is the radiation of particles accelerated at the shock front. Both hadronic and leptonic components are generally modeled with a power-law momentum distribution: $f_i(p) = k_i \left( \frac{p}{p_{\text{inj}}} \right)^{-\alpha}$, where $i = p$ (protons) or $i = e$ (electrons), with a spectral index $\alpha$ and injection momentum $p_{\text{inj}}$. The normalization $k_i$ is set by the CR acceleration efficiency $\xi_{\text{CR}}$. See Equations 9–11 in \cite{cardillo2016supernova} for further details.

But to explain the UHE emission we expect both acceleration and adiabatic compression of the crushed clouds to come into play as explained in \cite{blandford1982radio}. 
We define the compression factor as:
\begin{align}
s \equiv \frac{n_{\text{cc}}}{n_d} = \frac{n_{\text{cc}}}{r_{\text{sh}} n_0}
\end{align}
where $n_0$ the ambient density and $r_{sh} = \frac{n_d}{n_0}$ is the compression ratio at the shock. Given $B_0 = b \sqrt{n_0/\text{cm}^{-3}}\, \mu\text{G}$, the increase in the compressed magnetic field and from it the density of the crushed cloud are estimated as:
\begin{equation}
\frac{B_{cc}^2}{8 \pi} = n_0\mu_H v_{\text{sh}}^2 \, \rightarrow \, n_{cc} \simeq 94\, \left(\frac{n_0}{\text{cm}^{-3}}\right)^{3/2}\, \left(\frac{B_0}{\mu\,\text{G}}\right)^{-1}\, \left(\frac{v_{\text{sh}}}{10^7\,\text{cm/s}}\right)
\end{equation}
The resulting particle spectrum then shifts due to compression: $f^{\prime}(p) = f(s^{-1/3} p)$

However, as noted by Sushch and Borse \cite{sushch2023limits}, previous models may overestimate the amount of material in the compressed cloud. The maximum number of particles is limited by the shock-swept volume: $N_{\text{max}} = V_{\text{SNR}}\, n_0$. Assuming a volume filling factor $\chi$, the crushed cloud volume will be $ V_{\text{cc}} = \chi V_{\text{SNR}}$, leading to a total particle number $N_{\text{cc}} = \chi\, V_{\text{SNR}}\, n_{\text{cc}}$, which imposes the constraint:
\begin{equation}
    \chi \,s \leq 1
\end{equation}

In practice, due to limited interaction surface and time, $\chi s \ll 1$, and a more realistic upper limit is $\chi s \leq 0.1$. Although their analysis excludes cases consistent with previous models \cite{tutone2021multiple}, we adopt it here for a conservative treatment.

\subsection{Model Fitting}

\begin{figure}[htbp]
  \centering
  \includegraphics[width=0.49\textwidth]{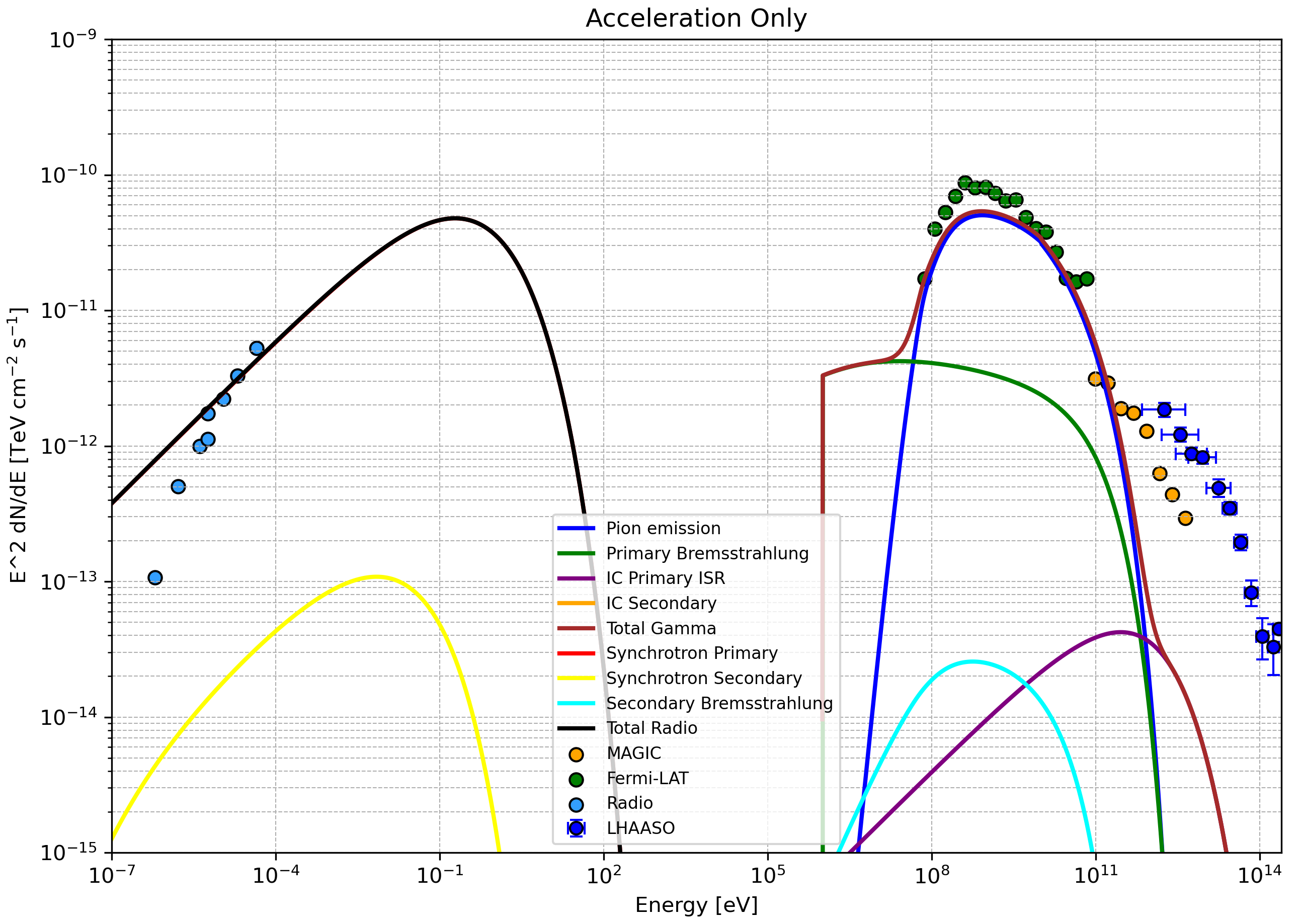}
  \hfill
  \includegraphics[width=0.49\textwidth]{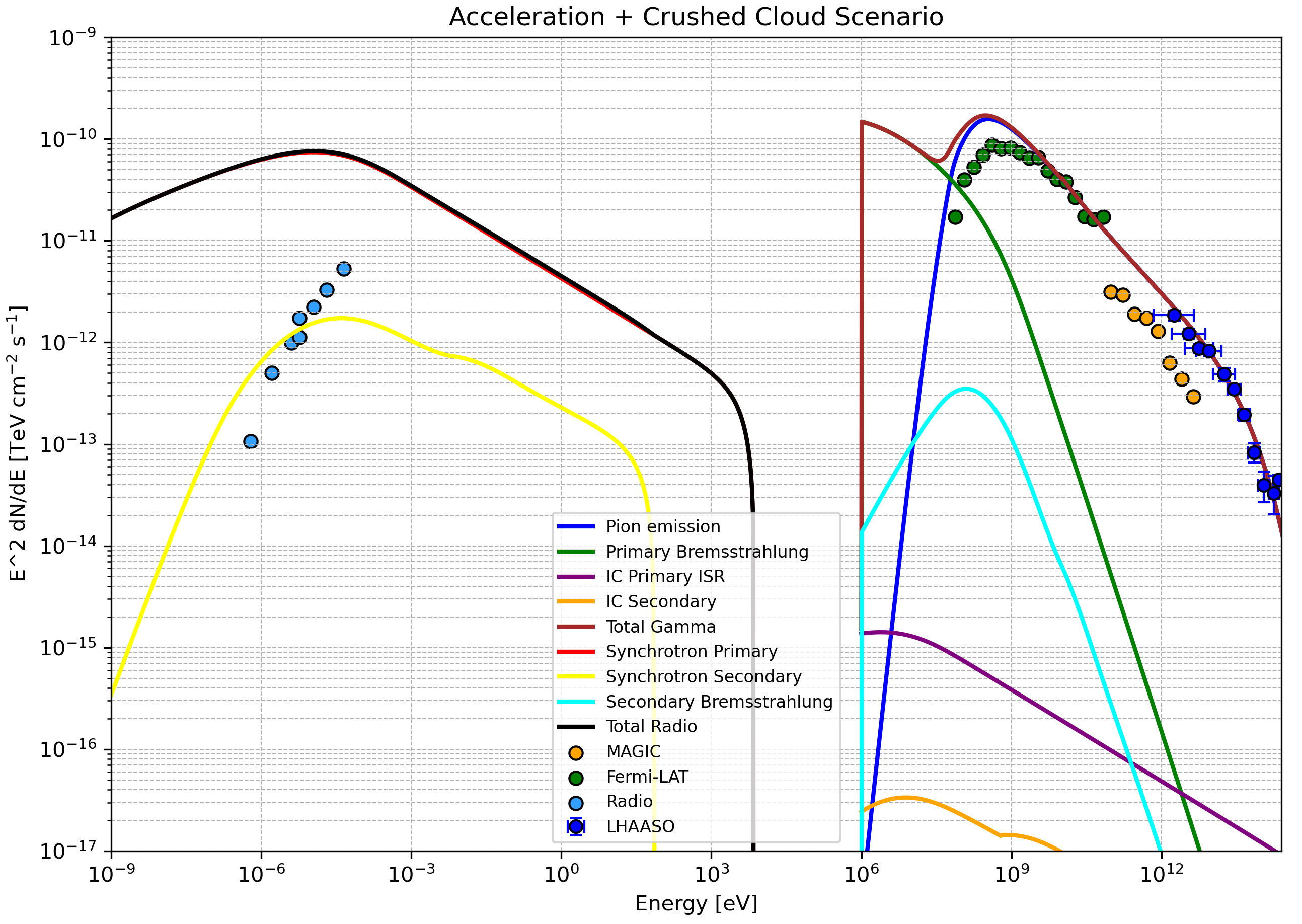}
  \caption{The figure illustrates the emission spectrum from the two scenarios analyzed along with observations in radio \cite{dae1994thermal} and gamma \cite{jogler2016revealing, aleksic2012morphological, cao2024evidence}. \emph{Left}: Depicts the scenario of acceleration at the shock front moving into the clouds. This model successfully fits certain radio data points and the HE data from Fermi, considering a hard spectral index of 4.2 and an ambient magnetic field of \(B_0 = 86\,\mu\mathrm{G}\). However, it does not reproduce the UHE data points, as anticipated. \emph{Right}: Shows the scenario of acceleration + clouds being crushed by the shock. This model successfully fits HE and UHE gamma data points, while there is an over estimation for the radio data. We used a spectral index of 4.6, as reported by LHAASO, an ambient magnetic field of \(B_0 \sim 180\,\mu\mathrm{G}\), and an interaction time of 3600 years.}
  \label{fig:Emission}
\end{figure}

Based on observational and theoretical estimates, we fix several parameters of the SNR in our calculations: the radius is set to $R_{SN} = 24\, \text{pc}$ \cite{cao2024evidence}, the shock velocity to $v_{sh} = 480-490\, \text{km/s}$ \cite{koo1995rosat}, and the ambient density to $n_0 = 20 - 25 \, \text{cm}^{-3}$ \cite{fang2010non}. Although there is some debate in the literature regarding the age and distance of W51, we adopt the values from \cite{tian2013high, zhang2017disentangling}, taking the distance as $d = 4.3\, \text{kpc}$ and the age as $t_{age} = 18,000$ years. We assume the acceleration efficiency of the SNR to lie between 6\% and 7\%, and following \cite{sushch2023limits}, we use a filling factor of 0.01 (1\%). Our approach follows the methodology outlined in \cite{cardillo2019orion, cardillo2016supernova}, and assuming a Kolmogorov perturbation spectrum with $ k_T = 2/3$, we derive the emission spectrum shown in Figure \ref{fig:Emission}. 

From figure \ref{fig:Emission} we can see that the acceleration scenario, focusing on particles accelerated from the shock, could effectively account for the radio data. On the other hand, incorporating the crushed cloud scenario alongside acceleration, and applying more stringent parameters, offers a compelling explanation for the observed HE-UHE $\gamma-$rays. However, the theoretical possibility of a single model accounting for both the radio and gamma-ray observations remains under investigation, including scenarios involving particle re-acceleration \cite{cardillo2016supernova}. Additionally, further analysis is needed to clarify the contribution of the proposed MC illumination mechanism.

\section{Conclusion}\label{Sec:Conclusion}
In this work, we have shown the possible perspectives and potential for CTAO and the ASTRI Mini-Array in the study of the morphology of SNR-MC interaction regions with the example of W51C-B, with two observed signatures of particle acceleration - \textit{pion bump} and an unprecedented upper limit of $\sim 200$ TeV $\gamma$-rays. Using the python library \emph{Gammapy}, we simulated 200 and 300 hours of source observation for CTAO and ASTRI Mini-Array respectively, exploiting recent LHAASO results with an aim to get more information on the morphology of the source. We assumed that the source emission must be detected from the SNR and the MC where the interaction happens - both hadronic in nature. Our morphological analysis confirms that with sufficient hours of observation CTAO and ASTRI can resolve the UHE emission region.

Our modeling suggests that the acceleration scenario, where particles are energized by a shock propagating into molecular clouds, could account for the radio data. The UHE $\gamma-$ray emission can be explained with combined contribution from acceleration and the adiabatic compression scenario. Both our modeling and simulations suggest a potentially larger population of particles below 10 TeV, on the contrary to what was observed by MAGIC. The potential for a unified theoretical model to simultaneously explain both the radio and $\gamma$-ray data is still being explored. This includes consideration of re-acceleration processes and  the MC illumination scenario.

\begingroup 
\footnotesize 
\textbf{Acknowledgment:} AS is funded under PNRR - CTA+ PROGRAM (Proposal: IR0000012) by the European Union - NextGenerationEU and approved by the MUR following Public Notice No. 3264 of 28/12/2021.
\endgroup 

\begingroup 
\footnotesize 
\bibliographystyle{unsrt}
\bibliography{references}
\endgroup

\end{document}